# Anomalous thermoelectric transport of Dirac particles in graphene

Peng Wei, Wenzhong Bao, Yong Pu, Chun Ning Lau, and Jing Shi

Department of Physics and Astronomy, University of California, Riverside, CA 92521

We report a thermoelectric study of graphene in both zero and applied magnetic fields. As a direct consequence of the linear dispersion of massless particles, we find that the Seebeck coefficient $S_{xx}$ diverges with $1/\sqrt{|n_{2D}|}$, where $n_{2D}$ is the carrier density. We observe a very large Nernst signal $S_{xy}$ (~ 50 μV/K at 8 T) at the Dirac point, and an oscillatory dependence of both $S_{xx}$ and $S_{xy}$ on $n_{2D}$ at low temperatures. Our results underscore the anomalous thermoelectric transport in graphene, which may be used as a highly sensitive probe for impurity bands near the Dirac point.

The unusual band structure of graphene gives rise to a host of intriguing phenomena in electrical transport properties that have been under extensive experimental investigations.[1-6] In solids, both charge and heat flows are simultaneously generated when an electrochemical potential or a temperature gradient is present, leading to additional effects. Fundamentally related to the electrical conductivity, other transport coefficients such as thermal conductivity and thermoelectric coefficients are also determined by the band structure and scattering mechanisms. Thermoelectric coefficients in particular, involve the energy derivatives of the electrical transport counterparts such as the conductivity $\sigma$ and the Hall angle $\Theta_H$. The anomalies in the latter are very often amplified and cause markedly distinct features in the former near the Dirac point. Furthermore, in the regime where the Mott relation is applicable, the relationship between the measured electrical conductivity and the Seebeck coefficient reveals how the chemical potential depends on the gate voltage or carrier density, which is dictated by the energy dispersion. Therefore, the thermoelectric transport coefficients can offer unique information and are complimentary to the electrical transport coefficients. A number of theoretical predictions have been made on transport coefficients other than electrical conductivities in graphene[7,8,9] which to date remain experimentally unexplored.

Single layer graphene sheets are mechanically exfoliated onto degenerately doped silicon substrates that are covered with 300 nm of silicon oxide. After locating suitable graphene sheets, we perform standard electron-beam lithography to attach electrodes in Hall-bar geometry. The electrodes consist of 7 nm of Cr and 100 nm of Au, and also serve as local thermometers. A micro-fabricated heater located on the right of *Fig. 1a* generates nearly parallel constant temperature contours along the graphene sample. The thermal emf generated is measured across the two parallel Cr/Au electrodes ~ 20 μm apart. These also double as local thermometers whose resistance is measured by the four-point method. A temperature difference of ~10 mK between the two Cr/Au wires can be readily measured for temperature $T$ >10 K. An additional pair of Cr/Au leads is used for transverse (Hall or Nernst) voltage measurements. All measurements were carried out in a cryostat with $T$ ranging from 1.5 to 300 K and magnetic field $\boldsymbol{B}$ up to 8 T. In this work, the results are from two representative devices (#1 and #2) out of approximately two dozen fabricated devices. They are single-layer graphene as determined from optical images, and are often corroborated by the well-defined half-integer quantum Hall effect at low temperature. The carrier mobility $\mu_c$ is typically ~3,000 cm$^2$/Vs.

We generate a temperature gradient and measure both the temperature difference $\Delta T$ and thermal emf change $\Delta V_{th}$. Fitting a straight line to the $\Delta V_{th}$ vs. $\Delta T$ data, we extract the Seebeck coefficient $S_{xx} = -\frac{\Delta V_{th}}{\Delta T}$ from the slope (*Fig. 1b*).[10] At zero magnetic field, $\sigma$ exhibits the characteristic minimum at $V_g \sim V_D$, the Dirac point. *Fig. 2a* shows $V_{th}$ as a function of $V_g$ for three temperatures. $V_{th}$ undergoes a sign change at the Dirac point $V_g = V_D$ = 10 V, indicating the carrier type changes from hole to electron as $V_g - V_D$ is swept from negative to positive. $V_{th}$ has a finite slope near $V_D$ over a 20 V range in $V_g$ which corresponds to ~ ±100 meV change in chemical potential $\mu$ measured from the Dirac point. This region coincides with the minimum in $\sigma$, where charged impurities modify the conductivity.[11-14] As $V_g$ is further away from $V_D$ on both sides, the magnitude of $V_{th}$ decreases, scaling approximately with $1/\sqrt{|V_g - V_D|}$ (dashed line in *Fig. 2a*). This $V_g$ dependence is more noticeable in the linear dependence of $1/V_{th}^2$ on $V_g$ (*Fig. 2b*). The solid lines are the power-law fits with exponent ~ 0.95 and cross zero in the vicinity of the Dirac point from both sides, indicating a diverging behavior of $S_{xx}$. Note that near the Dirac point, $V_{th}$ crosses zero, and the $1/\sqrt{|V_g - V_D|}$-dependence breaks down, as denoted by the hatched region. For comparison, the same $V_{th}$ data is also plotted as $1/|V_{th}|$ vs. $V_g$ in *Fig. 2c* and the straight lines are drawn in the linear region. Clearly, the $1/V_{th}^2$ plot shows a better linear relationship with $V_g$ over the whole range. In addition, $1/V_{th}^2$ extrapolates to zero at almost the same $V_g$ for different temperatures, but $1/|V_{th}|$ does not.

The fact that $|V_{th}|$ or $|S_{xx}|$ diverges as $1/\sqrt{|V_g - V_D|}$ is actually a direct manifestation of the linear dispersion of the Dirac particles in graphene. Let us assume $\sigma \sim |\mu|^\alpha$, which is sufficiently general to include both dirty ($\alpha \sim 2$) and clean ($\alpha \sim 1$) limits.[12,13] For degenerate electron systems, we expect the Mott relation $S_{xx} = -\frac{\pi^2 k_B^2 T}{3e} \frac{\partial \ln \sigma(\mu)}{\partial \mu}$ to hold, yielding $S_{xx} \sim -\frac{1}{\mu}$ for highly doped regimes. On the other hand, for a 2D system with a linear dispersion relation, then we expect $\mu = \hbar v_F \sqrt{n_{2D} \pi} \propto \pm \sqrt{|V_g - V_D|}$, where the +(-) sign corresponds to the electron- (hole-) doped regime, and $v_F$ is the Fermi velocity. Combining these relations, we have $S_{xx} \sim \frac{-\text{sgn}(\mu)}{\sqrt{|V_g - V_D|}}$. This is in contrast to the ordinary 2D electron systems with a quadratic dispersion relation, in which $\mu \propto n_{2D}$, and hence $S_{xx} \sim \frac{-1}{V_g - V_D}$. From this diverging behavior of $S_{xx}$, we can conclude that the dispersion relation is linear rather than quadratic, as expected for Dirac particles. It is worth noting that the exponent $\alpha$ is absorbed in the pre-factor of $S_{xx}$ and does not affect the functional dependence of $S_{xx}$, as is the case in $\sigma$. This makes the thermoelectric transport uniquely sensitive to the electronic band structure.

Not every device shows the electron-hole symmetry shown in *Fig. 2*. *Fig. 3a* displays $S_{xx}$ vs. $V_g$ of a different device with $V_D \sim 33V$ for several values of $T$ ranging from 11 to 255 K. Away from $V_D$ on the hole side, $S_{xx}$ decreases with decreasing $V_g$, similar to the behavior of the previous device. In contrast, $S_{xx}$ stays flat on the electron side, indicating a strong electron-hole asymmetry as seen in $\sigma$ by others.[12] Near $V_D$, we observe a broad transition region in $S_{xx}$ connecting the electron- to hole-doped regimes. Furthermore, $S_{xx}$ follows different $T$-dependence for different $V_g$ (in *Fig. 3b*). Near $V_D$, the magnitude of $S_{xx}$ is close to zero. Away from $V_D$ on the hole side, e.g. at $V_g = 0$ V or $\sim 33$ V left of $V_D$, $S_{xx}$ is nearly a straight line for the whole temperature range. As $V_g$ approaches $V_D$ from the hole side, $S_{xx}$ begins to deviate from the linear $T$-dependence at progressively lower temperatures. On the electron side, however, even at $V_g = 60$ V (or $\sim 30$ V right of $V_D$), $S_{xx}$ remains non-linear in $T$ except at very low temperatures.

The departure from the linear $T$-dependence is an indication of the potential breakdown of the Mott relation. For this device, when $|V_g - V_D| = 30$ V, $|\mu|$ is about 160 meV measured from the Dirac point. It is reasonable to expect high-order corrections in the Sommerfeld expansion at relatively high temperature where the condition $|\mu| \gg k_B T$ fails. For graphene, another relevant energy scale is the bandwidth $\gamma$ of impurity states[15,16] near the Dirac point. The Mott relation only holds if $\frac{\gamma}{k_B T} \gg 1$, which ensures $\sigma$ to be a slow-varying function of energy over this band of impurity states.[16] In the impurity scattering model, this band can be highly asymmetric due to the finite scattering potential. Here we attribute the departure from the linear $T$-dependence on the electron side to the asymmetric nature of the band of impurity states. For this reason we only focus on the relatively low-T region on the hole side where the Mott relation apparently holds. Since $S_{xx}$ is proportional to $\alpha T$, and inversely proportional to $\mu$ or $v_F\sqrt{|n_{2D}|}$, we plot $S_{xx} \cdot \sqrt{|n_{2D}|}$ (called $\beta$) vs. T in the inset of *Fig. 3b*. Extracted from the slope, $v_F$ ranges from 0.8 to 1.6 x10$^6$ m/s depending on the value of $\alpha$ (from 1 to 2), which is in good agreement with the values obtained by others.[17] In relating $V_g$ to $n_{2D}$ for above estimations, we use $n_{2D} = \dfrac{C_g V_g}{e} + \bar{n} = \dfrac{C_g}{e}(V_g - V_D)$, where $C_g$ is the capacitance per unit area and $\bar{n}$ is the induced density by charged impurities at the Dirac point. A value of $C_g$ = 103 aF/μm$^2$ is determined from our Hall data.

In a magnetic field, carriers diffusing under $\nabla T$ experience the Lorentz force, resulting in a non-zero transverse voltage $V_y$. The transverse effect or the Nernst effect is measured by $S_{xy} = -\dfrac{E_y}{|\nabla T|} = \dfrac{\Delta V_y}{\Delta T_x}$. In non-magnetic metals, $S_{xy}$ is negligibly small (~10 nV/K per tesla).[18] In ferromagnets, spin-orbit coupling can lead to a large spontaneous Nernst signal.[19] Here we observe an exceedingly large Nernst peak (~ 50 μV/K at 8 T) at the Dirac point (Fig. 4a), and we attribute it to the unique band structure of graphene. In classical transport, the Mott relation takes the following form:[7,20]

$S_{xy} = -\dfrac{\pi^2}{3}\dfrac{k_B^2 T}{e}\left(\dfrac{\partial \Theta_H}{\partial \varepsilon}\right)_\mu = \dfrac{\pi^2 k_B^2 TB}{3}\dfrac{\partial}{\partial \mu}\left(\dfrac{\tau}{m^*}\text{sgn}(\mu)\right)$. $S_{xy}$ is directly proportional to the energy derivative of the Hall angle $\Theta_H$ or inversely proportional to the cyclotron mass $m^*$. For massless particles, the vanishing cyclotron mass can indeed lead to a diverging behavior in $S_{xy}$. In graphene devices, however, the anomaly is diminished by the impurity states near the Dirac point. Recall that the Mott relation breaks down in this region. Here we estimate the magnitude of $S_{xy}$ at the Dirac point both from $\Theta_H$ outside this region where the Mott relation holds and from $\gamma$. Since we have $\Theta_H = -\mu_c \cdot B \cdot \text{sgn}(\mu)$ ($\mu_c$:

carrier mobility), we obtain $\Delta\Theta_H \sim 2.2$ with an 8T magnetic field at 255 K. This change in $\Theta_H$ occurs over $\gamma \sim$204 meV as estimated from the width of the conductance minimum, yielding $S_{xy} \sim$ 68 µV/K. This is in very good agreement with the experimentally observed peak value (~ 50 µV/K). Additionally, $\Theta_H$ is directly proportional to **B**, which indicates a linear **B**-field dependence in $S_{xy}$, with an estimated slope of ~5.4 µV/K*T at 160 K. Indeed, the linear **B**-dependence of $S_{xy}$ is observed (*Fig. 4a*), and the slope of the straight line is ~ 6 µV/K*T. Similar to $S_{xx}$ whose diverging behavior is greatly modified by the disorders, the anomaly in $S_{xy}$ depends on the carrier mobility as well as $\gamma$. We expect to see more pronounced anomalous behavior in both $S_{xx}$ and $S_{xy}$ in cleaner samples.

At low temperatures and *B*=8 T, the device conductance exhibits clear quantum Hall plateaus as $V_g$ is varied. In this regime, we observe oscillations in $S_{xx}$ (*Fig. 4b* and *4c*) that are reminiscent of the Shubnikov-de Hass oscillations in $\rho_{xx}$,[2,3] and the side peaks and dips in $S_{xy}$ that correlate with the oscillatory structures in $S_{xx}$. At *T* = 11 K, $S_{xx}$ shows peaks (dips) as $\mu$ is inside the broadened Landau levels (LL) on the hole (electron) side. These peaks (dips) correspond to the LL indices *n* = 1 and *n* = 2 for holes (electrons). $S_{xy}$ also changes sign at these fillings. It is also worth noting that $S_{xx}$ crosses zero at the Dirac point (in the lowest LL), accompanied by an additional small dip (peak) on the hole (electron) side. The origin of this feature is unknown, but it could reveal some peculiarities of the zero-th LL at high magnetic fields. In conventional 2D electron systems, the observed $S_{xx}$ peaks at the LL's are consistent with the calculations in the integer quantum Hall regime.[21] In graphene samples, the *n* =1 and *n* =2 peaks in $S_{xx}$ on both electron and hole sides are also expected. However, we do not observe vanishing $S_{xx}$ as $\mu$ is located between the two adjacent LL's. The non-vanishing $S_{xx}$ was previously attributed to the activated behavior in ordinary 2D electron systems. In our samples, the relatively large magnitude of $S_{xx}$ between the LL's may be caused by the broadened LL's due to disorders. We expect to see $S_{xx} \rightarrow 0$ at low temperatures and the predicted activated behavior at high temperatures in cleaner samples.

As the temperature increases, the oscillations in $S_{xx}$ and $S_{xy}$ become weaker, although the overall magnitude of both $S_{xx}$ and the central peak in $S_{xy}$ increases (*Fig. 4c*). As discussed earlier, the characteristic width of the Nernst peak is primarily determined by $\gamma$ which is greater than $k_B T$. The Nernst width remains nearly unchanged as a consequence.

In summary, the diverging behavior ($|S_{xx}| \sim 1/\sqrt{|n_{2D}|}$) of the Seebeck coefficient along with the exceedingly large Nernst peak at the Dirac point is characteristic of the massless particles in graphene. With disorders, these generic anomalies are somewhat masked near the Dirac point. However, the diverging behavior can be retrieved from those quantities as the chemical potential approaches the Dirac point. In higher mobility graphene samples, the anomalies are expected to be more drastically pronounced.

Authors would like to thank Chandra Varma for suggesting the thermopower study on graphene, and thank Chris Dames, Vivek Aji, Shan-Wen Tsai, and Vicent Ugarte for many helpful discussions. PW, YP and JS acknowledge the support of DOE DE-FG02-07ER46351 and ONR/DMEA H94003-07-2-0703. WB and CNL acknowledge the support of NSF CAREER DMR/0748910, NSF CBET/ 0756359 and ONR/DMEA Award H94003-07-2-0703.

**Note Added**: During the preparation of this manuscript, we became aware of related work with a similar conclusion from Zuev et al.[22]

Figure Captions:

***Fig.1 (a)*** SEM image and circuit schematic of a graphene device for thermoelectric measurements. **(b)** $\Delta T$ vs. thermo-voltage $V_{th}$ for a series of heater power steps at 255K and zero gate-voltage. The linear fit of this curve gives the thermopower of 39 µV/K.

***Fig.2 (a)*** $V_{th}$ vs. $V_g$ for three different temperatures. The 16K data (red circle) was multiplied by a factor of five. The dash lines are the fits described by $|S_{xx}| \sim 1/\sqrt{|V_g - V_D|}$.
**(b)** $1/V_{th}^2$ vs. $V_g$ plot for the same data shown in (a). The shaded area is for $|V_g - V_D| < 10$ V. Red lines are the best power-law fits with exponent ~ 0.95. **(c)** $1/|V_{th}|$ vs. $V_g$ plot for the same data in (a). Red lines are straight lines as guides to the eye.

***Fig.3 (a)*** $V_g$-dependence of longitudinal Seebeck coefficient $S_{xx}$ at different temperatures (11K – 255K) and zero magnetic field. **(b)** $T$-dependence of $S_{xx}$ at different gate-voltages. The inset is the $T$-dependence of $\beta = S_{xx}\sqrt{|n_{2D}|}$ at $V_g$=0 V for low temperatures. The slope of the linear fit is proportional to $\alpha/v_F$.

***Fig.4 (a)*** $V_g$-dependence of Nernst signal $S_{xy}$ at 160 K with different magnetic fields (1 – 8 T). Inset: B-dependence of $S_{xy}$ at $V_g = V_D$, and the red line is a linear fit. **(b)** Two-terminal conductance $G$ and thermopower $S_{xx}$ vs. carrier density $n_{2D}$ at $T$ = 11 K and **B** = 8 T. The corresponding Landau level index $n$ is shown on the top axis. **(c)** $S_{xx}$ (black triangle) and $S_{xy}$ (red circle) vs. Landau level index $n$ for four different temperatures at **B** = 8 T.

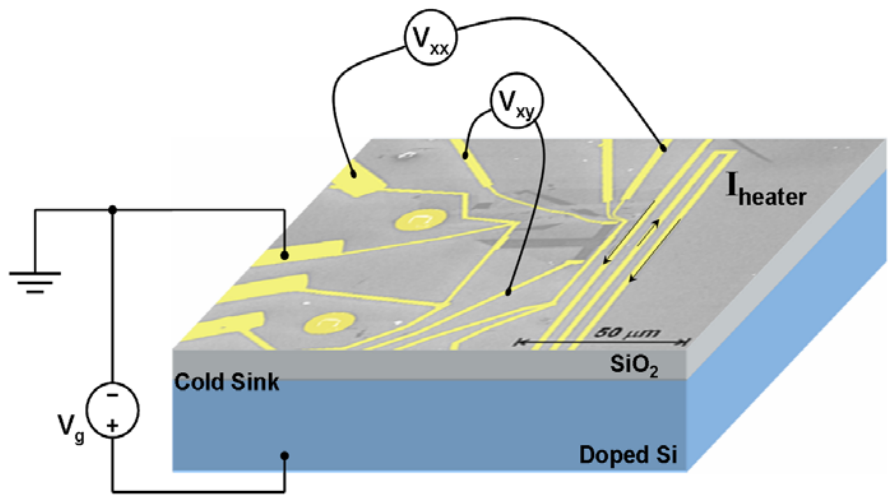

(a)

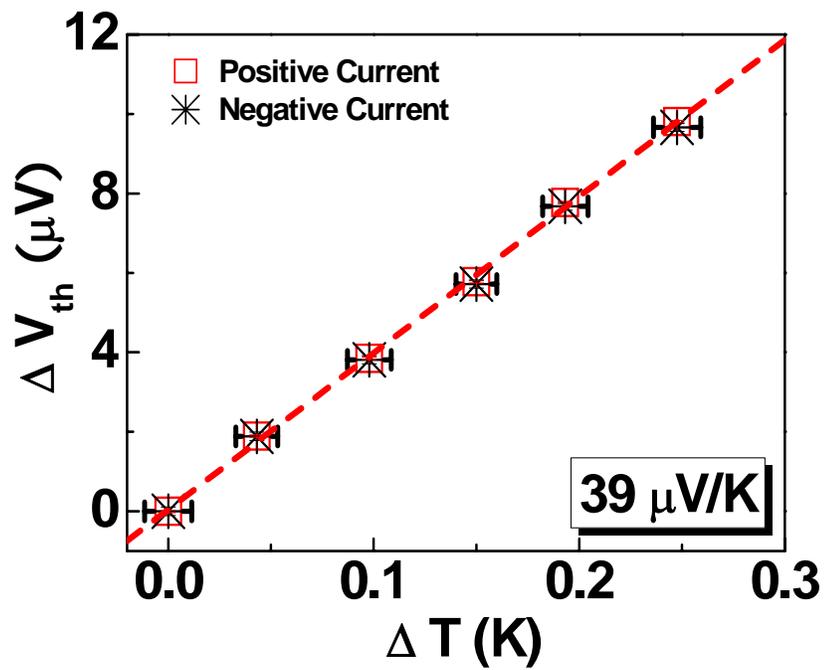

(b)

Figure 1

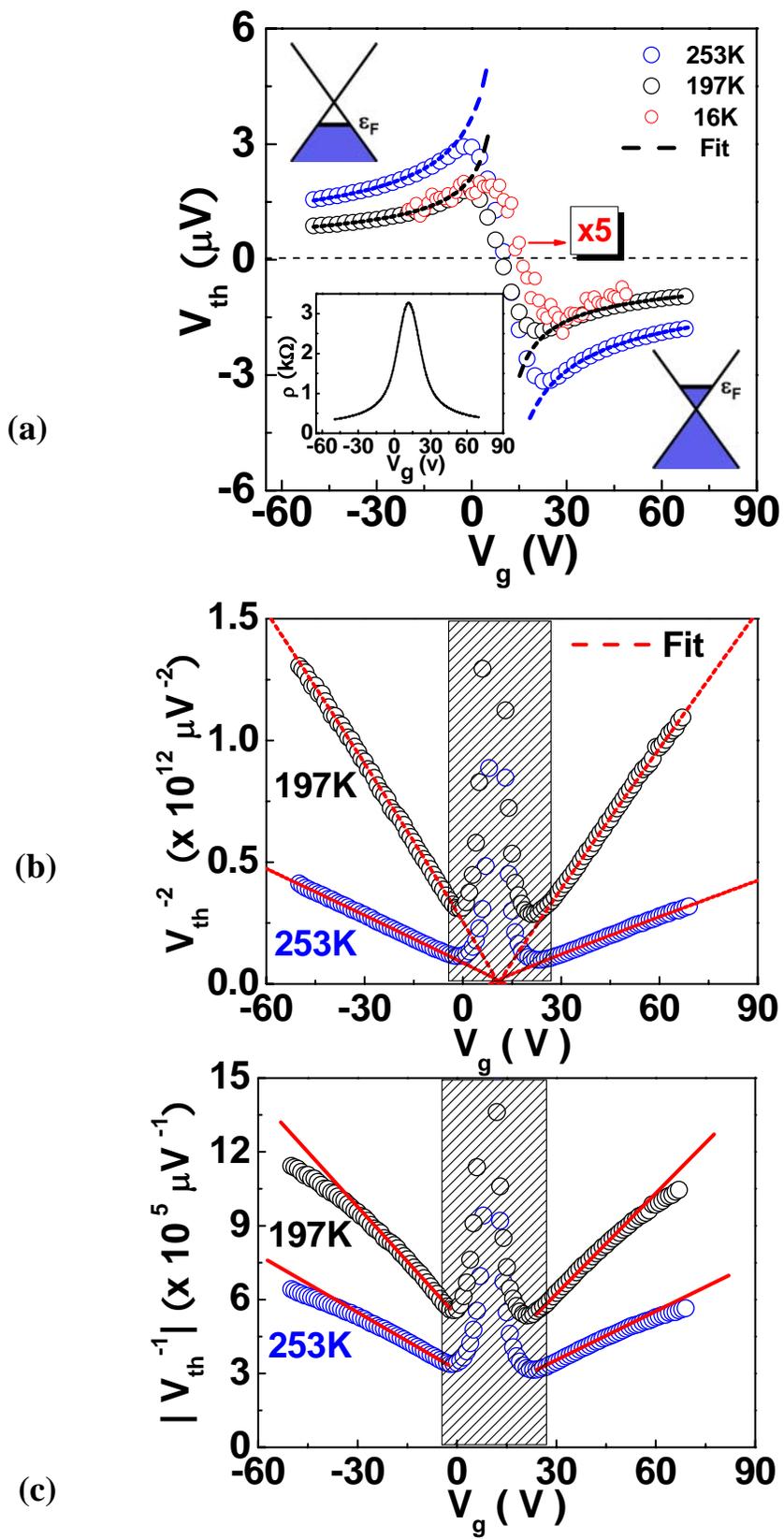

Figure 2

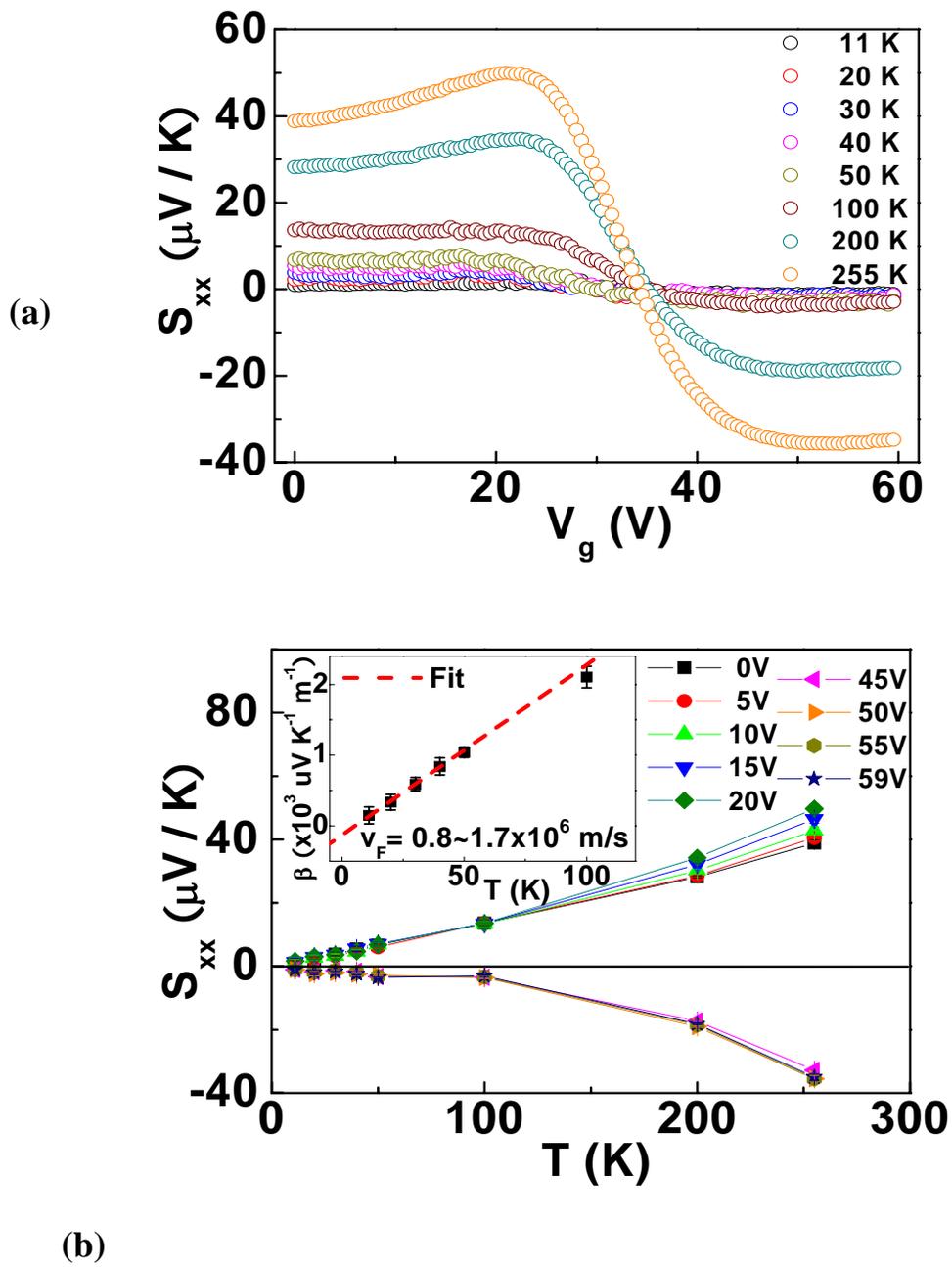

Figure 3

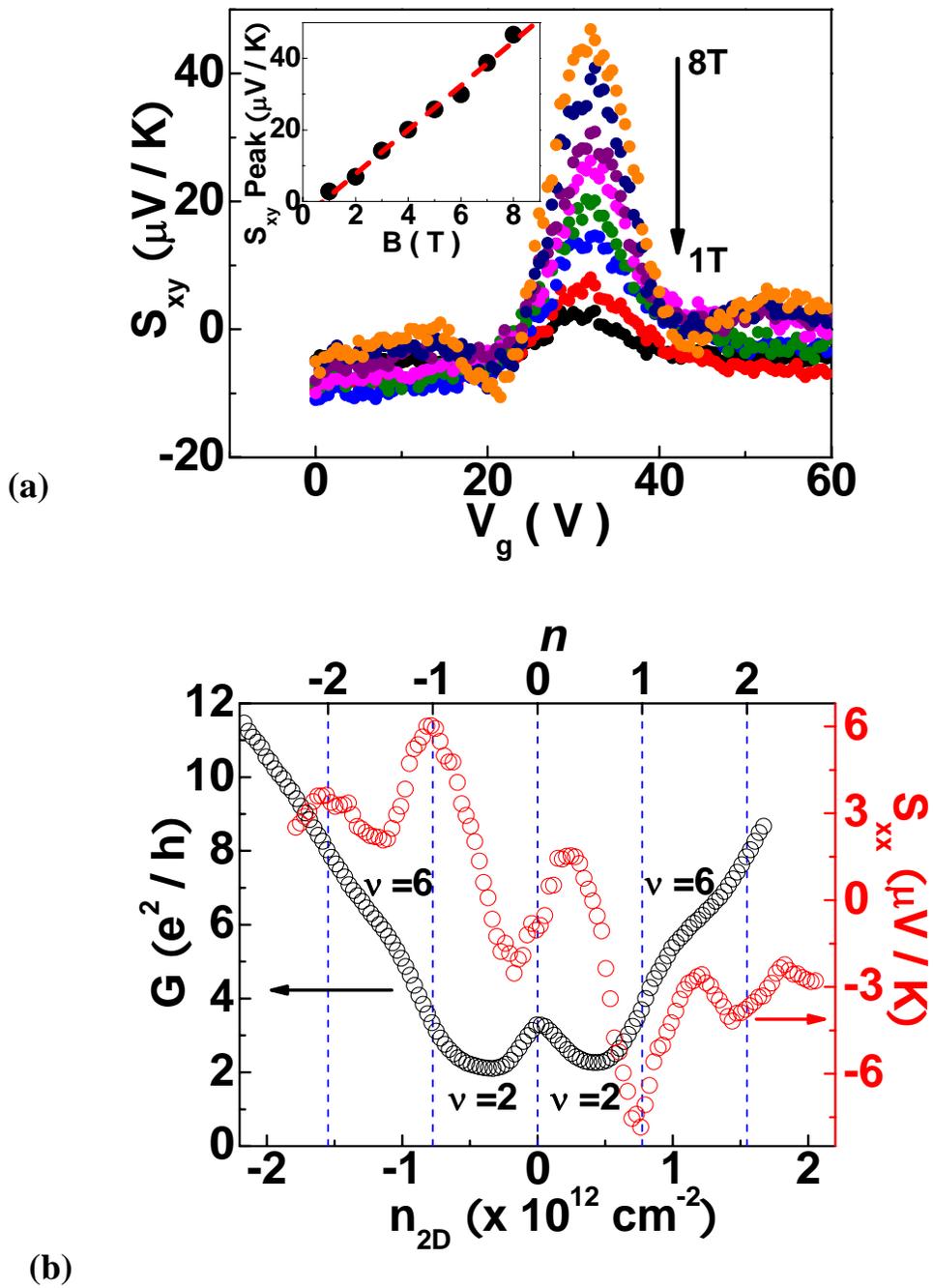

Figure 4a & 4b

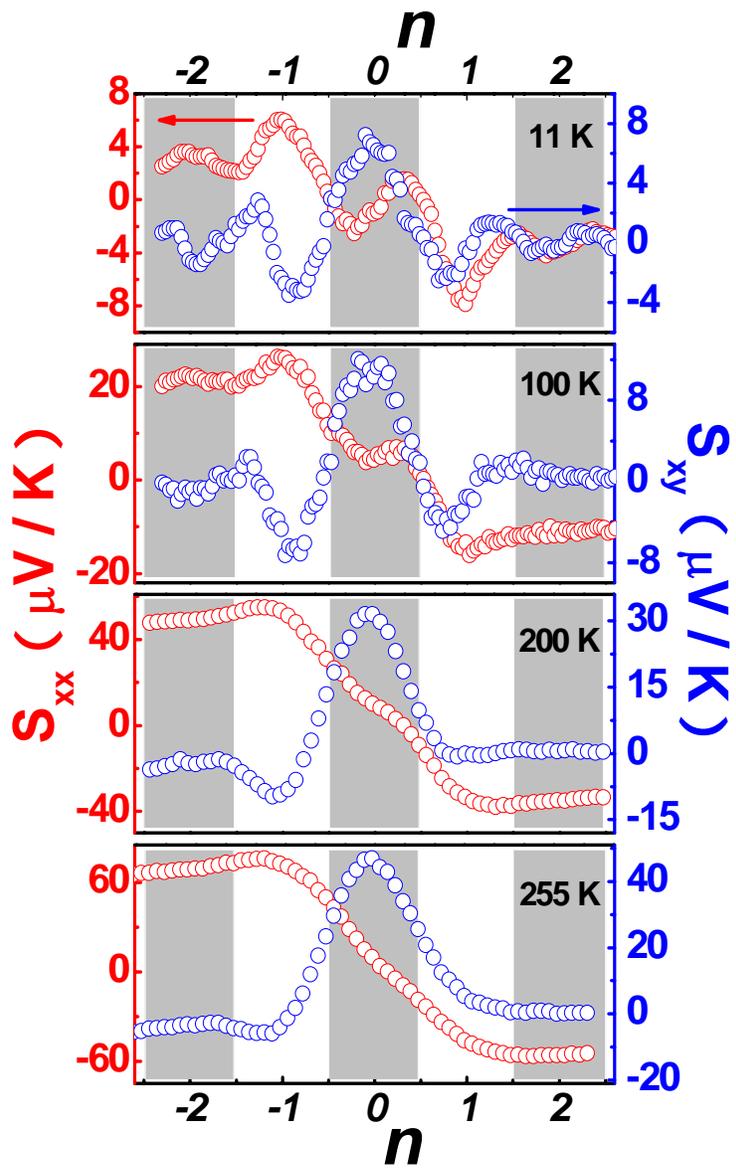

Figure 4c